# Computational comparison of a novel cavity absorber for parabolic trough solar concentrators

**Khaled Mohamad, Philippe Ferrer**
School of Physics, University of the Witwatersrand, Johannesburg 2001, South Africa

khaled@aims.edu.gh

**Abstract** Parabolic trough mirror plants are a popular design for conversion of solar energy to electricity via thermal processes. The absorber of concentrated solar radiation can reach high temperatures (> 500°C) and is responsible for efficiency losses mainly via thermal radiation. We build on our previous work on hot mirrors to study an absorber with a mirrored cavity. The cavity absorber for the parabolic trough receiver is designed to trap solar and thermal radiations by reflecting them back onto the absorber very efficiently, which would otherwise be lost. This paper shows simulation results indicating that the proposed design can exceed the heat transfer fluid temperature compared to existing alternatives. Using a theoretical model we developed, we can infer the temperature profile for the receiver unit, from which efficiency parameters can be derived.

1. Introduction

A parabolic trough solar thermal power plant consists of a series of parabolic mirrors concentrating solar radiation onto a linear focal line along which the receiver unit is positioned. The receiver heats up and in turn imparts a large portion of its heat to a heat transfer fluid circulating within. This heat transfer fluid can then be utilized in a steam cycle to generate electricity. The receiver is one of the most complex parts and the efficiency of the whole system largely depends on it. It has to be carefully designed in such a way so as to minimize the energy losses. Every part of the receiver unit is a topic of ongoing research, such as the working fluid that can be used, and also the optical, chemical, and thermal properties of the concerned material [1].

Typically, the receiver unit consists of a blackened absorber pipe (AP) encapsulated by the glass cover (GC), (See Fig. (1b)). There is a vacuum in between to minimize convective losses [2]. For the conduction losses, the thermal contacts between the receiver pipe and the glass cover are kept to a minimum. The heat transfer fluid (HTF) inside the receiver pipe is heated by the concentrated solar radiation. The hot HTF can be used in generating electricity through a steam cycle or in thermochemical applications [3]. The dominant heat losses at high temperatures are due to the thermal emission (IR) from the receiver pipe. There is a conventional method to minimize the IR by painting the receiver pipe with a spectrally selective coating, a dielectric film that absorbs well in the visible region of the solar spectrum and emits poorly in the IR region. Much work has been published in regard to the selective coating and their properties [4]. The main weakness of selective coating is that it prevents the receiver pipe from being heated to high temperature, since it thermally decomposes at about 680 K [5][6].

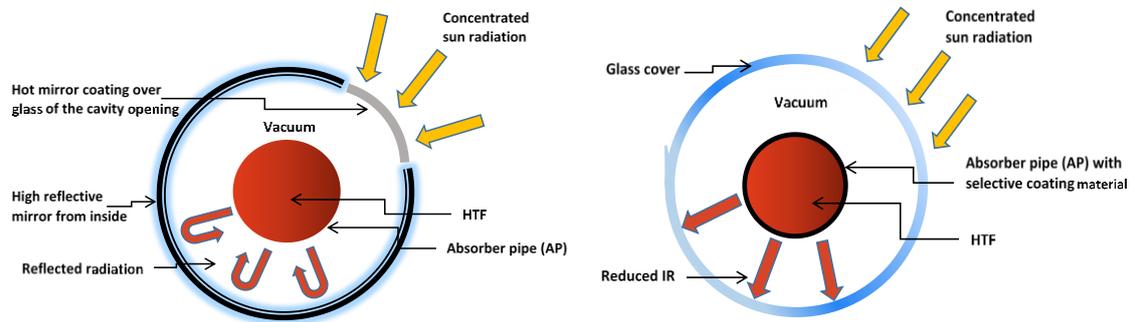

**Figure 1.** a) Receiver unit with the cavity design.　　b) Receiver unit with a selective coating.

An alternative option to the selective coating, which this work aims to discuss, is to reduce the thermal emission from the glass cover tube of a trough collector by trapping IR via a reflective surface on the part of the glass not facing the trough. The solar radiation inlet may be coated with a hot mirror type coating. The new aspect of this paper is applying the reflective cavity around the absorber, which is shown in Fig. (1a).
It consists of borosilicate glass over the opening, which is coated with a hot mirror film and the remaining circumference is a high reflective aluminium mirror.

Hot mirror coating films have been an active area of research in many applications, seeking to improve efficiency and reduce heat radiation losses [6,7]. It is often utilized in applications related to energy conservation and protection purposes, i.e., light bulb envelop, furnace windows, welding and laser goggles, and astronaut helmets [7]. The hot mirror coating for a solar collector must meet some performance specifications. It needs to be highly transparent (> 90 %) in the visible region and have high reflectivity in the IR region of the solar spectrum. There are two general types of hot mirror films: a semiconducting oxide with a high doping level and a very thin metal film sandwiched between two dielectric layers (see [7,8] for more details). The coating with a thin metal film shows some unavoidable losses. On the other hand, the semiconducting oxide with a high doping level shows more advantages, i.e., Indium-Tin-Oxide (ITO).

## 2. Theory and simulation study

We briefly review the total heat transfer of the system and the interaction between its components. The physical basis of the model that we are using starts with a complete description of the thermal interaction, which is shown in Fig. (2) and then applying energy conservation principles for the thermal interactions between the components of the receiver [9].

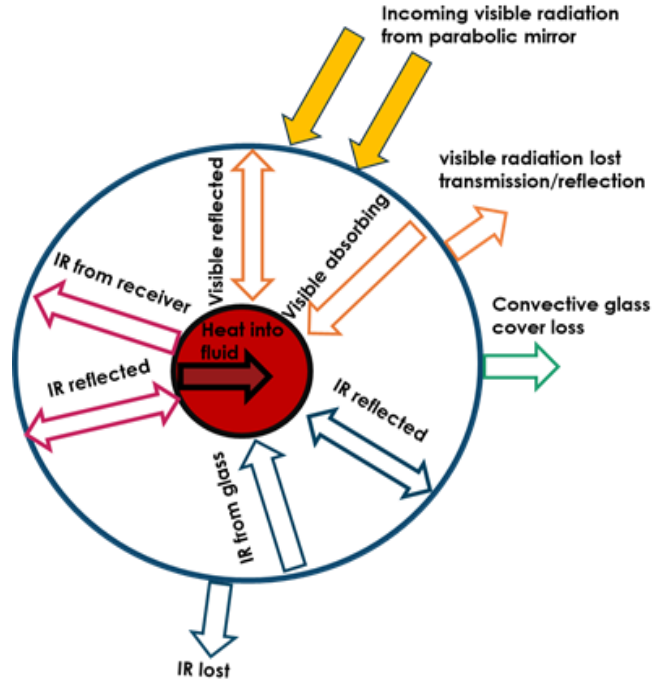

**Figure 2.** Schematic representation of the possible heat transfer modes.

The net heat flux due to solar radiation $q'_{Sol}$, convection $q'_{Conv}$, radiation $q'_{Rad}$, and conduction $q'_{Cond}$ are computed under steady state conditions using the energy balance relationship

$$\left(\sum q'_{Sol} + \sum q'_{Conv} + \sum q'_{Cond} + \sum q'_{Rad}\right)_{ij} = 0. \qquad (1)$$

The AP, the GC, and the HTF are discretized into control volumes (CV), using a finite volume method (FVM). The AP and GC are discretized along azimuthal Fig. (3b) and longitudinal directions Fig. (3a), but HTF is only discretized along the longitudinal direction Fig. (3a). Eq. (**Error! Reference source not found.**) holds for every CV.

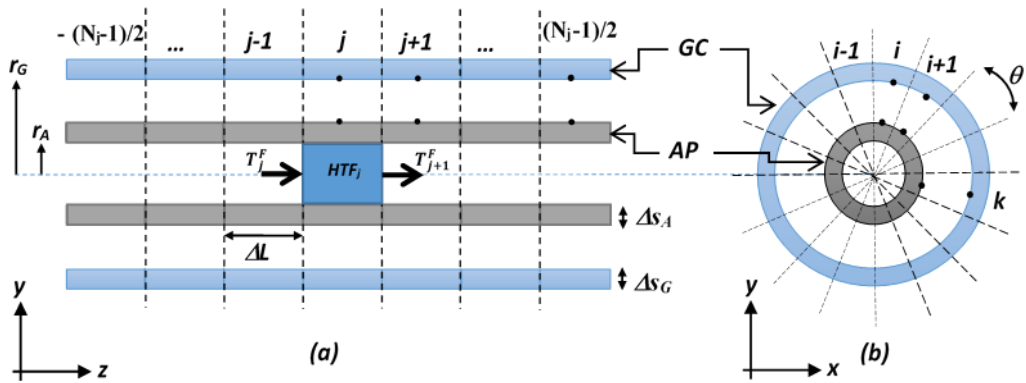

**Figure 3.** The discretization of AP and GC into control volumes [9].

The detailed theoretical calculations including the numerical solutions that have been implemented in a code are discussed in [9].

## 3. Results

The theoretical model of the solar receiver, taking hot mirror interactions into account, has been derived in [9]. For the purposes of this research, the simulation code presented in [9] is edited in order to fulfil the cavity requirements. A simulation validation for the work done in this paper was undertaken using two approaches. We selected simulation parameters for physical scenarios where the outcomes could be derived by other means. We used the zero irradiation case, zero conductivity in the materials, zero HTF convective coefficient, and zero emissivity on the absorber surface. The results conformed to theoretical expectations. Secondly, upon comparing the simulation results with existing experimental data for a selective coating, we found that the comparison was encouragingly close (less than 0.7% discrepancy), see [9].

The operating conditions and design parameters that were used simulate the SEGS LS2, the LS-2 is one of three generations of parabolic troughs installed in the nine SEGS (Solar Electric Generating System) power plants in California, [10]. They are shown in table 1.

**Table 1.** Design parameters of the SEGS LS2 used in our simulation [10]. * stands for the cavity design requirements.

| Parameter | Value |
| --- | --- |
| Collector aperture (W) | 5 m |
| Focal distance (f) | 1.84 m |
| Absorber internal diameter | 0.066 m |
| Absorber external diameter | 0.07 m |
| Absorber emissivity (IR) | 0.15 |
| Glass internal diameter | 0.109 m |
| Glass external diameter | 0.115 m |
| Glass emissivity (IR) | 0.86 |
| Receiver absorptance (visible) | 0.96 |
| Glass transmittance (visible) | 0.93 |
| Parabola specular reflectance | 0.93 |
| Incident angle | 0.0 |
| Solar irradiance | 933.7 W/$m^2$ |
| HTF | Molten salt |
| Mass flow rate | 0.68 Kg/s |
| Temperature HTF (inlet) | 375.35 K |
| Temperature ambient | 294.35 K |
| Wind speed | 2.6 m/s |
| Reflectivity of the cavity mirror* | 0.95 |
| ITO reflectivity (IR)* | 0.85 |
| ITO transmittance (visible)* | 0.875 |

The minimum cavity opening size is related to an upper limit to the possible concentration ratio of a parabolic mirror on its own, which is related to focal line width [11]. On the basis of that, the arc length of the minimum cavity opening is 4.8 cm for the design parameter in table 1.
In Figures (4, 5, 6), the results of the simulated temperature profiles for the AP, GC, and HTF are shown. In Fig. (4), the axial temperature variation for the HTF along the receiver is displayed, for HTF inlet temperatures of 375 K that enters the absorber at one end, and as a result of the flow rate of 0.68 kg/s along the length L (m) of the absorber tube under the concentrated solar irradiance, the HTF heats up. HTF temperature increases roughly linearly and then flattens out to approach the stagnation temperature (where solar energy input equals IR losses). The maximum HTF temperature for this design rises close to 1370 K.

Figure (5) shows the temperature profile around the AP circumference. The angle 180° points directly away from the sun (and towards the centre of the parabolic mirror). The temperature varies by approx. 200 K around the circumference. Three profiles are shown, each 50 m further along the AP. The temperature increase along the axial direction is relatively high. In order to make this simulation more realistic, finite sun-size effects will still be included. These will alter the solar radiation profile incident on the absorber pipe, and likely result in a more even temperature spread.

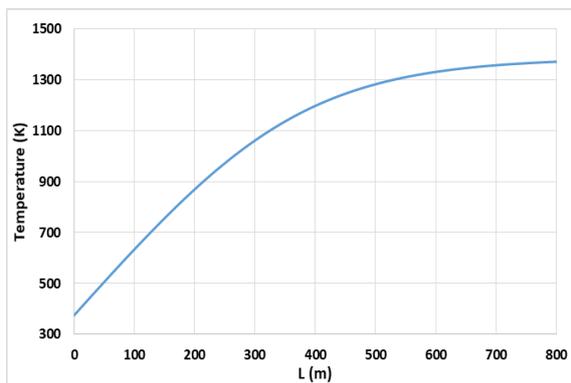

**Figure 4.** HTF temperature distribution for 375 K inlet temperature.

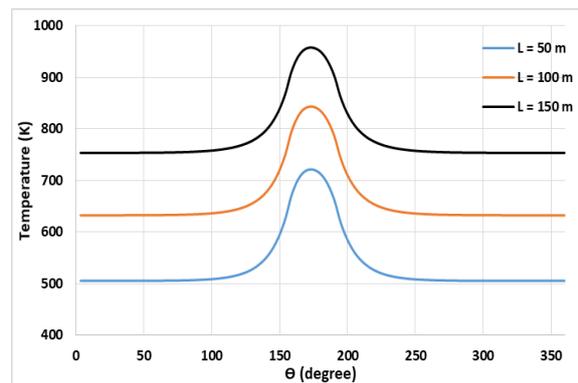

**Figure 5.** Temperature around AP circumference taken at different distances.

Figure (6) displays the temperature profile around the GC circumference. The temperature around the circumference varies by approx. 320 K and the temperature along the axial direction is very small. Hence most thermal losses will occur from the radiation entrance window.

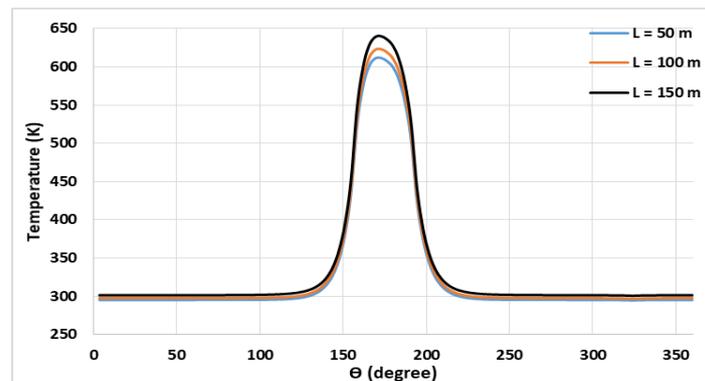

**Figure 6.** Temperature around GC circumference taken at different distances.

We evaluate the performance of the cavity design by comparing it to alternatives with bare and selective coating.
Figure (7) display the temperature profiles around the AP at 150 m. The temperature of the part of the AP surface that faces the parabolic mirror for the cavity design is higher than the selective coating. This is due to the more concentrated radiation incident on the AP.
The GC in Fig. (8) at 150 m, the part of the GC surface facing the parabolic mirror for the different designs indicates a higher temperature for the cavity design compared to the selective coating and bare.

At the remaining circumference, the cavity design has a much lower temperature than the other designs along the entire length.

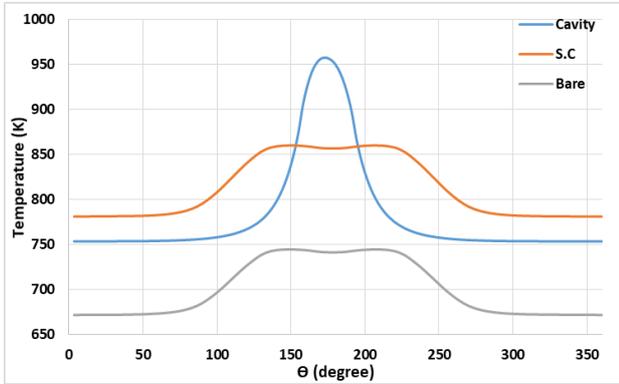

**Figure 7.** Temperature around AP circumference taken at 150 m receiver length for different designs.

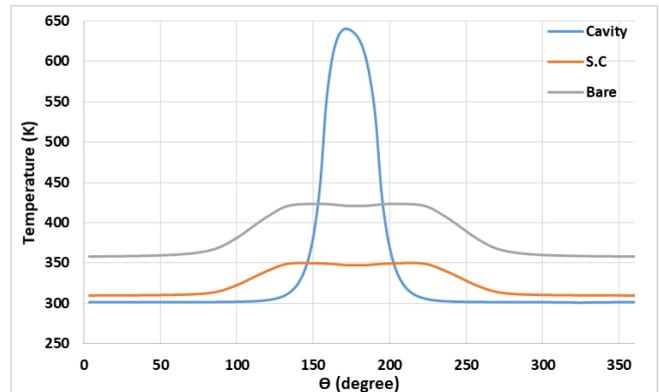

**Figure 8.** Temperature around GC circumference taken at 150 m receiver length for different designs.

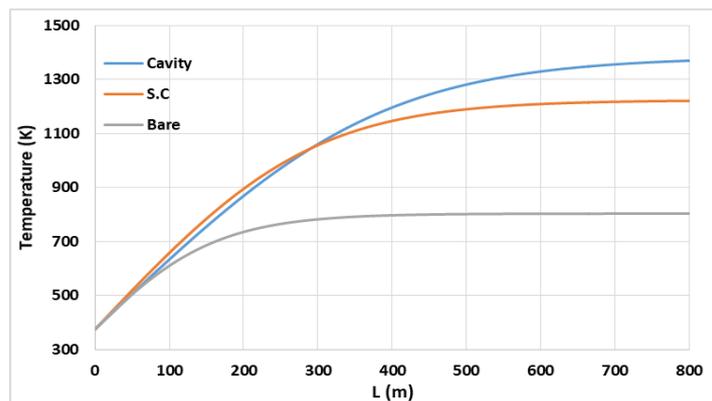

**Figure 9.** HTF outlet temp for 375 K inlet temperature of different designs.

Figure 9 shows the HTF temperature along the length, which is close to the surface temperature of the AP. The selective coating material on the AP will be chemically decomposed around 680 K, as previously mentioned and its length will be limited to < 100 m.

The HTF temperatures along the receiver unit in the cavity is capable of exceeding the selective coating temperature ceiling, as shown in Fig. (9).

## 4. Conclusion

We introduced a cavity concept to reduce thermal radiation losses for receivers for parabolic trough solar plants, and compared it to existing systems. The cavity design performs very well at higher temperatures and it theoretically capable of exceeding 1300 K, thus outperforming current technologies due to its thermal stability. This in turn can increase the overall (Rankine) efficiency of the entire plant. There are further important parameters that affect the temperature profile and the efficiency, such as the cavity opening size and the reflectivity of the cavity mirror. These will be shown in further communication.